\documentstyle[twoside, epsfig]{article}

\input ibvs2.sty

\begin{document}
\IBVShead{6168}{25 April 2016}

\IBVStitle{BG CMi time keeping}

\IBVSauth{Michel Bonnardeau$^1$}

\IBVSinst{MBCAA Observatory, Le Pavillon, 38930 Lalley, France, email: arzelier1@free.fr}

\SIMBADobjAlias{BG CMi}
\IBVStyp{GCVS}
\IBVSkey{photometry}
\IBVSabs{Twelve seasons, from 2005 to 2016, of photometric monitoring of the intermediate polar BG CMi are presented and are compared with previous observations. The spin up of the white dwarf is found to have undergone a major change between 1996 and 2005. There is also some evidence for sideband signals.  }

\begintext
BG Canis Minoris (RA=07h 31min 29.00s, DEC=+09$^{\circ}$ 56' 23.1", J2000) is an intermediate polar (a cataclysmic system in which the white dwarf is magnetized enough to module the accretion). Its magnitude is around 14.5. The orbital motion has a period  $P_{orb}=3.235$ hr, and gives a modulation visible by photometry. 
\\ \\
There is also a modulation of the light curves with a period $P_{spin}=913$ s. This modulation is usually interpreted as being due to the spin of the white dwarf. However this is not firmly established: the spin period may be twice $P_{spin}$, with both poles visible (Patterson \& Thomas, 1993), or the observed modulation may be synodic, with the spin period being actually shorter (Norton et al, 1992). 
\\ \\
According to Pych et al, 1996 and references therein, the period $P_{spin}$ is decreasing, so there is a spin-up of the white dwarf. However, according to Kim et al, 2005 (hereafter K05), the rate of this spin-up is decreasing.
\IBVSfig{6.0cm}{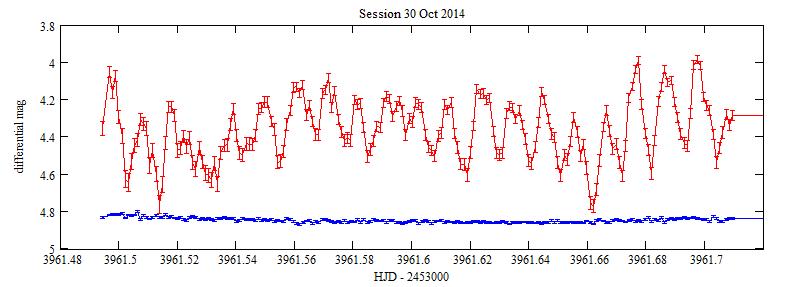}{Upper light curve: BG CMi, lower one: the check star shifted by +4.2 mag. The error bars are the quadratic sum of the 1-sigma statistical uncertainties on the variable/check star and on the comparison star. }

\textbf{}
\\
\textbf{Observations.} Photometry observations of BG CMi were carried out over twelve seasons, from 2005 to 2016, with a 203 mm f/6.3 Schmidt-Cassegrain telescope, a clear filter and an SBIG ST7E camera (KAF401E CCD). The exposures were 60 s long.  For the aperture differential photometry, the comparison star is GSC 768-01373. 5161 useful images were obtained over 49 nights. A check star, GSC 768-01665, is used to compare the standard deviations to the statistical uncertainties so as to make sure that the systematic errors are low. An example of a light curve is given in Figure 1. 
\\ \\
\\
\textbf{O-C analysis of the $P_{spin}$ modulation.} The light curves are searched for pulses due to the $P_{spin}$ modulation. $N=254$ well defined pulses are found.
These pulses are then fitted with the $t(E)=T+P.E+B.E^2$ quadratic ephemeris, where E is the cycle number, minimizing S: 
\\
$S=\sqrt{{ \frac{\sum\limits_{i=0}^{N-1}(t_{i}-T-PE_{i}-BE_{i}^2)^2/\delta t_{i}}{\sum\limits_{i=0}^{N-1}1/\delta t_{i}}}}$
\\
where the $t_{i}$ are the HJD of the pulses and the $\delta t_{i}$ their uncertainties.
\\ \\
Solution(s) are searched using a Monte Carlo algorithm. This has the advantage over the least squares method of readily revealing cycle counting ambiguities. It works the following way:
\\ \\
make 10 runs (or more) of the following: 
\begin{itemize}
\item make 1 millions trials (or more) of the following: 
\begin{itemize}
\item select randomly a set of T, P, B;
\item compute the cycle number $E_{i}$ of each pulse;
\item compute S;
\end{itemize}
\item retain the set of T, P, B that gives the smallest S.
\end{itemize}
\medskip
The sets of T, P, B are randomly selected in the following ranges: \\
$T=$ the first pulse $\pm P_{spin}/2$ \\
$P=P_{spin} \pm P_{spin}/1000$ \\
$B=-20.10^{-13} \pm 20.10^{-13}$
\\ \\
For each run, the same number of cycles is always obtained, 375,969, between the first pulse and the last pulse. So there is no cycle ambiguity. The adopted values for T, P, B are the average values of the runs, and
the adopted values for the uncertainties are the standard deviations:
\\ 
 \\
\hspace*{0.5cm}$T = 2453449.38837(48)$ $HJD$ $= 2453449.389105$ $BJD_{TDB}$
\\
\hspace*{0.5cm}$P = 0.0105726046(47)$ $d$          \hspace{9cm}         (1)
\\
\hspace*{0.5cm}$B = -1.41(11)10^{-13}$ $d$
\\ \\
The white dwarf is then spinning up with $\dot{P}=2B/P=-2.67*10^{-11}$ in a time scale $P/2|\dot{P}|=543$ kyr.  
\\ \\
The ephemeris (1) is precise enough to be applied with no cycle ambiguity to the 2002-2005 data of K05. The cycle numbers run from -76,998 to -4,194.
\\ \\
The spin-up rate in the ephemeris (1) is quite smaller than the one obtained from 1982-1996 observations (Pych et al, 1996), 
$t(e)=T_{P96}+P_{P96}.e+B_{P96}.e^2$, with \\ \\
\hspace*{0.5cm}$T_{P96}=2445020.2800(2)$ $HJD$ $=2445020.2806$ $BJD_{TDB}$
\\
\hspace*{0.5cm}$P_{P96}=0.010572992(2)$ $d$  \hspace{9cm} (2)
\\
\hspace*{0.5cm}$B_{P96}=-3.83(4)10^{-13}$ $d$
\\ \\
When the ephemeris (2) is applied to the 2002-2005 data of K05, the first pulse is at number 720,254, with an ambiguity of $\pm 1$ or more cycles. 
\\ \\Besides the heliocentric correction, there are smaller corrections to be taken into account (Eastman et al, 2010), in particular the leap seconds due to the Earth rotation slowing down. The leap second correction at the time of the ephemeris (1), in 2005, is 32 s, and 36 s in 2016. And at the time of ephemeris (2), in 1982, it is 20 s, and 30 s in 1996. The barycentric effect of Jupiter and Saturn is neglected as it is only $\pm 4$ s and cyclic (unlike the leap seconds that keep accumulating), and the other general relativistic corrections are much smaller. (And there is also 32.184 s to be added to obtain $BJD_{TDB}$.)
\\ \\
This gives the O-C diagram of Figure 2, computed from the ephemeris (1).
\IBVSfig{10.5cm}{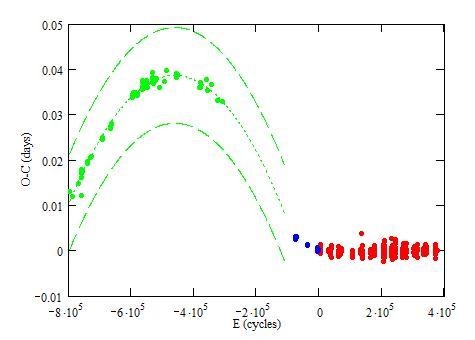}{Red dots: present data, Blue: K05, Green: Pych et al (1996) with the first pulse at $E_{0}=-720,254-76,998$.
Green dotted line: ephemeris (2) with $E = e+E_{0}$, upper dash line: with $E = e+E_{0}+1$, lower dash line: with $E = e+E_{0}-1$.}
\textbf{}
\\
Between 1996 and 2005 there was a change of regime, from the ephemeris (2) to the ephemeris (1). Unfortunately, this was not observed, except by K05 who captured the end of this episode. 
\\ \\
\\
\textbf{Fourier analysis of the orbital modulation.} The light curves are inspected to look for the orbital dips, and they are found in phase with the orbital ephemeris of K05. 
\\ \\
The light curves are analyzed with the PERIOD04 software program (Lenz \& Breger, 2005), which provides simultaneously sine-wave fitting and least-squares fitting algorithms, around the frequency $1/P_{orb}$. Besides $1/P_{orb}$, up to 6 harmonics are used to fit the orbital modulation. This yields the orbital period:
\\
$P_{orb}=0.134,748,376(74)$ $d$
\\
which is in agreement with the period of K05 (the uncertainty is given by the PERIOD04 Monte Carlo simulation).
\\ \\
\\
\textbf{Fourier analysis of the residuals.}
As the $P_{spin}$ period is varying, the data are analyzed season by season (so the variation is not too important): for each season an average $P_{spin}$ period is computed from the ephemeris (1) and the PERIOD04 program is used to derived the amplitude and phase. The same is done for 2 harmonics (actually, the amplitudes of these 2 harmonics are very small: the modulation is quasi-sinusoidal). 
\\ \\
The data are then prewhiten with this fit for the $P_{spin}$ modulation and with the fit for the orbital modulation, yielding a Fourier spectrum of the residuals for the season.
\\ \\
These Fourier spectra show many variations from one season to the other. But it is not clear which are physical and which are numerical accidents. Actually, such features were already observed (Patterson \& Thomas, 1993, Garlick et al, 1994, de Martino et al, 1995). All these spectra for each season are summed, resulting in a spectrum for the whole set of observations,
prewhiten with the $P_{spin}$ and the orbital modulations. Most of the variations cancel out, and  a fairly strong peak shows up with two fainter ones, as shown in Figure 3.
\\
\\
The peak (1) corresponds to a period of 1083 s. It was already observed by Patterson \& Thomas, 1993 who interpreted it as the sideband $1/P_{spin}-2/P_{orb}$.
\\ \\
The peak (2) corresponds to a period of 1309 s. It has not been reported by other observers. It could be  the sideband $1/P_{spin}-4/P_{orb}$.  
\\ \\
The peak (3) corresponds to a period of 835 s. This is close to the signal at 847 s reported by Norton et al, 1992 and Choi et al, 2007 in X-ray, but unseen in optics except by Garlick et al, 1994, and which was interpreted as the sideband $1/P_{spin}+1/P_{orb}$.
\\ \\ \\
\IBVSfig{9cm}{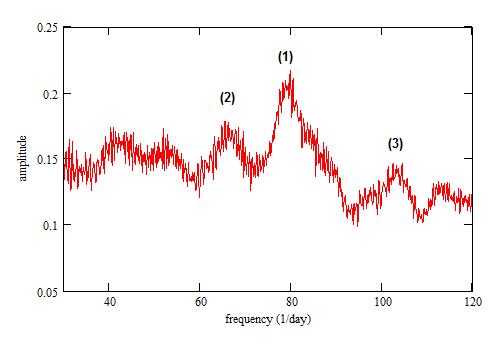}{Sum of the spectra of the residuals for the twelve seasons.} 
\references
Choi C.-S., Dotani T., Kim Y., Ryu D., 2007, \textit{New Astronomy}, \textbf{12}, 622. 

de Martino D. et al, 1995, \textit{A\&A}, \textbf{298}, 849.

Eastman J., Siverd R., Gaudi B.S., 2010, \textit{PASP}, \textbf{122}, 935.

Garlick M.A. et al, 1994, \textit{MNRAS}, \textbf{267}, 1095.

Kim Y.G., Andronov I.L., Park S.S., Jeon Y.B., 2005, \textit{A\&A}, \textbf{441}, 663.

Lenz P., Breger M., 2005, \textit{Comm. Asteroseismology}, \textbf{146}, 53.

Norton A.J., McHardy I.M., Lehto H.J., Watson M.G., 1992, \textit{MNRAS}, \textbf{258}, 697.

Patterson J., Thomas G., 1993, \textit{PASP}, \textbf{105}, 59.

Pych W., Semeniuk I., Olech A., Ruszkowski M., 1996, \textit{Acta Astronomica}, \textbf{46}, 279.
\endreferences
\end{document}